\documentclass[12pt,a4paper]{article}
\usepackage{geometry}
\usepackage[utf8]{inputenc}
\usepackage{mathtools}
\usepackage[T1,T2A]{fontenc}
\usepackage[english]{babel}
\usepackage{amsmath,amssymb,amsfonts,amsthm}
\usepackage{paratype}
\usepackage{graphicx}
\usepackage{cite}
\usepackage{xcolor}
\usepackage{comment}

\usepackage{cite}
\usepackage[linktocpage=true, colorlinks=false, linkcolor = blue, citecolor = red]{hyperref}
\begin{document}
	\title{Global isometric embedding\\ of the Kerr spacetime and beyond}
	
	\author{ Mikhail Markov\thanks{ Physique de l’Univers, Champs et Gravitation,
Université de Mons -- UMONS,
Place du Parc 20,
7000 Mons, Belgium, e-mail: \texttt{markovmikhailv@gmail.com}}, Anton Sheykin\thanks{Saint Petersburg State University, e-mail: \texttt{a.sheykin@spbu.ru}}}  
		\date{}
	\maketitle
	\begin{abstract}
		We construct a new isometric embedding of the Kerr spacetime into $(5+7)$-dimensional flat spacetime.  It is asymptotically flat, covers the entire $r>0$ region and has a very simple form.
		 We also show how this embedding can be generalized to any spacetime admitting a Kerr-Schild-like decomposition of the metric.
	\end{abstract}

\section{Introduction}

Relativists became interested in isometric embeddings into flat space shortly after the beginning of GR itself. The first embedding of the Schwarzschild spacetime was constructed in 1921 by Kasner \cite{kasner3}, but turned out to be quite useless as it did not allow one to pass through the horizon. The famous Fronsdal embedding \cite{frons} covered the entire manifold, but had another drawback, being not asymptotically flat, even though the metric itself was. This means that its surface did not reproduce a four-dimensional plane at large distances from the source. Therefore, the use of the Fronsdal embedding as a background for the perturbative analysis of small deviations from the Schwarzschild metric at large $r$ appeared to be problematic, which motivated the search for other embeddings.

The first asymptotically flat embedding of the Schwarzschild spacetime was found in 2010 \cite{blashke2010} and required $(2+5)$ ambient dimensions. In the following year, S. A. Paston developed a symmetry-based embedding method, which led to the construction of a $(1+5)$-dimensional asymptotically flat embedding of the Schwarzschild spacetime \cite{statja27}. 

Five years ago, in an attempt to embed the Kerr spacetime, we developed a modified version of the embedding construction algorithm. It exploits commuting one-parameter groups of isometries to reduce the embedding problem to the corresponding orbit space. Using it, we constructed a global embedding of a three-dimensional rotating black hole, but a four-dimensional rotating black hole remained a challenging case. 

The missing link was the idea that allowed Kerr to obtain the metric itself, namely, the Kerr-Schild decomposition \cite{Kerr:2009}. Using it, in this paper we construct an embedding \eqref{Kerr-embedding} of the Kerr spacetime into $(5+7)$-dimensional flat space. It has several appealing features:
\begin{itemize}
	\item it covers the entire $r>0$ region,
	\item it is asymptotically flat,
	\item  it is expressed in terms of elementary functions and has a very simple form.
\end{itemize}

Moreover, the embedding of the Kerr metric gave us an idea of how to embed similar spacetimes. In addition to the Kerr embedding, we present a local isometric embedding of any spacetime whose metric can be written in a Kerr–Schild-like form.
\section{The idea of embedding construction}\label{sec:idea}
The embedding method proposed in \cite{statja27} requires one to construct a representation of the symmetry group of the metric by some subgroup of the Poincaré group of the ambient spacetime. While it allows one, in principle, to obtain all possible types of embedded surfaces with given symmetry properties, the symmetry requirement is most effective in constraining the embedding ansatz when the metric has a sufficiently large symmetry group. For example, this method made it possible to classify all possible symmetric embeddings of the Schwarzschild spacetime into six dimensions; however, the absence of spherical symmetry in a rotating black hole greatly reduces the usefulness of this method in this case.

This method was modified in \cite{statja70} to avoid using the full group of isometries. The modification provides a way to reduce the problem of embedding a $(d+1)$-dimensional pseudo-Riemannian manifold to the problem of embedding a $d$-dimensional manifold using an Abelian subgroup of the isometry group. The core idea is to realize this symmetry locally\footnote{In this section, we do not discuss the global properties of the resulting embeddings, and all statements are strictly local. Nevertheless, the resulting embeddings may turn out to be global.} by ordinary (pseudo)rotations in the ambient spacetime. To do so, note that a choice of two functions $f,\psi\in C^\infty(M)$ and a parameter $\xi=\pm1$ defines a map from $M$ to $\mathbb{R}^2$, given by
\begin{align}\label{circle-like}
    y^1=f\cos\left(\sqrt{\xi}\psi\right)\,,\qquad y^2=\frac{f}{\sqrt{\xi}}\sin\left(\sqrt\xi\psi\right)\,,
\end{align}
where $y^1$, $y^2$ are the coordinates on $\mathbb{R}^2$, equipped with the flat metric 
\begin{align}
  \eta_2:=\pm\left((dy^1)^2+\xi(dy^2)^2\right)\,. 
\end{align}
Using this, the problem of isometric embedding of $(M,g)$ into $(\mathbb{R}^N=\mathbb{R}^2\oplus\mathbb{R}^{N-2},\eta_N=\eta_2\oplus\eta_{N-2})$ can be reduced to the problem of embedding $(M,g\mp(df^2+\xi f^2d\psi^2))$ into $(\mathbb{R}^{N-2},\eta_{N-2})$. Note that the residual tensor $g\mp(df^2+\xi f^2d\psi^2)$ may be degenerate, and the corresponding map need not be an embedding of $M$. Nevertheless, with a slight abuse of terminology, we will also use the term ``isometric embedding'' for maps inducing degenerate $(0,2)$ tensors.

Finally, as shown in \cite{statja70}, when the group of isometries contains several commuting one-parameter symmetries, we can iterate this procedure and choose $f$ and $\psi$ appropriately to reduce the problem to that of embedding the orbit space of $M$ under these symmetries, equipped with the residual tensor, into $\mathbb{R}^{N-2k}$ for some $k\in\mathbb{N}$. In practice, on a region where $V\neq0$, we choose coordinates $(x^m,\varphi)$ on $M$ in which the Killing vector associated with the symmetry is $V=\frac{\partial}{\partial\varphi}$ and choose the functions $f$, $\psi$ so as to eliminate all terms involving $d\varphi$ from the residual tensor while keeping its coefficients independent of $\varphi$. More specifically, in this case, $\psi$ is chosen to be affine in $\varphi$, while $f$ is chosen to be proportional to the square root of the absolute value of the metric component we aim to eliminate. Informally speaking, we use this procedure first to eliminate all off-diagonal $\varphi x^m$ components of the metric, and then the $\varphi\varphi$ component of the residual tensor. We emphasize that the residual tensor generally differs from the tensor obtained by orthogonally projecting the original metric onto the directions orthogonal to the symmetry orbits. This approach allowed us to construct an isometric embedding of the $(2+1)$-dimensional BTZ black hole into $7$-dimensional flat space in \cite{statja70}. 

Nevertheless, the presented algorithm has two subtleties. First, its direct application depends explicitly on the choice of coordinates $(x^m,\varphi)$. Second, after using all of the applicable symmetries of the metric, we still need to embed the residual tensor. For example, we have applied the algorithm to the Kerr spacetime in Boyer–Lindquist coordinates using its rotational and time-translation symmetries, but have not found a straightforward explicit embedding of the residual tensor on the resulting two-dimensional space. In fact, such an application leads to the Kuzeev embedding \cite{Kuzeev}, which contains two functions of $\theta$ and $r$ specified only implicitly.

In this work, we supplement the construction above with an additional ingredient that allows us to construct an embedding of the Kerr spacetime. Namely, we decompose the metric into a ``background'' $\bar g$ and a ``perturbation'' $h$\footnote{Note that we do not assume any linearization here. In fact, $h$ is more like a deformation than a perturbation, but we follow the terminology commonly used in the Kerr-Schild context.}:
\begin{align}\label{decomp}
    g=\bar g+h
\end{align}
in such a way that an isometric embedding of $\bar g$ is known (in this work, we take $\bar g$ to be either the Minkowski or the AdS metric). The problem then reduces to constructing a map inducing $h$; its direct sum with the background embedding gives an isometric embedding of $(M,g)$. The decomposition \eqref{decomp} is very general and includes, for example, the Kerr-Schild decomposition and some of its generalizations as special cases. In this work, we consider only cases in which $h$ has rank $1$, so locally $h_{\mu\nu}=Hl_\mu l_\nu$, but we impose none of the additional conditions often required in the Kerr-Schild setup.

\section{Examples: Schwarzschild and Kerr}
\subsection{Schwarzschild embedding}\label{sec:schw}
Let us examine the Schwarzschild metric as a toy model before turning to Kerr. The Kerr-Schild decomposition is manifest in Eddington---Finkelstein coordinates\footnote{Here $T=v-r$, where $v$ is the standard ingoing Eddington–Finkelstein null coordinate.}:
 \begin{align}\label{Schw}
	g=\eta-\frac{2M}{r}(d(T+r))^2\,,
\end{align}
where $\eta=dT^2-dr^2-r^2(d\theta^2+\sin^2\theta d\Phi^2)$.
We start with the trivial background embedding $y^{\mu} = x^{\mu}$, {$\mu=0,1,2,3$,} and then proceed to embed the KS term:
 \begin{align}\label{Schsincos}
 	y^4 =\sqrt{ \dfrac{2M^3}{r}} \cos \left(\frac{T+r}{M}\right)\,, \qquad y^5 =\sqrt{ \dfrac{2M^3}{r}} \sin \left(\frac{T+r}{M}\right).
 \end{align}
The metric induced by $y^0\,-\,y^5$ has an unwanted term $ -\left(\frac{d}{dr}\sqrt{\frac{2M^3}{r}}\right)^2=-\frac{M^3}{2r^3}$ in $g_{rr}$, so we need an additional (and timelike!) component to counter it:
 \begin{align}
	y^6 = \sqrt{\dfrac{2M^3}{r}}.
\end{align}
We therefore obtain an embedding in a flat space with signature $(+ - - - - - +)$ which was discovered in \cite{blashke2010}. Aside from its asymptotic flatness, it has an extremely simple form in comparison with all six types of six-dimensional embeddings listed in \cite{statja27} at the cost of an additional ambient dimension.

An important lesson from Schwarzschild is that the form of the metric most suitable for embedding is not necessarily the (most) diagonal one. 

\subsection{Kerr embedding}
To the best of our knowledge, there have been no successful attempts to construct an embedding of the full Kerr spacetime. A $(3+6)$-dimensional embedding of Kerr \cite{Kuzeev} and Kerr-Newman \cite{kuzeevRN} was found by Kuzeev, but it contained two functions $F_{1,2}(r,\theta)$ specified only implicitly and could not smoothly cover even the outer horizon.

However, the problem of constructing an embedding diagram, i.e. embedding only a Kerr submanifold (for example, the equatorial plane or the horizon) \cite{Sharp1981}, is considerably easier and has been studied quite extensively; see, e.g., \cite{Hong:2012fa}.

We start with the Kerr metric in the coordinates $(T,r,\theta,\Phi)$, in which it takes the form
\begin{align}\label{Kerr-metric}
	g=\eta-\frac{2Mr}{\Sigma}\left(d\left({T+r}\right)-a\sin^2\theta d\Phi\right)^2\,,
\end{align}
where $\Sigma=r^2+a^2\cos^2\theta$, $M>0$, and the Minkowski part is given by\footnote{These coordinates on Minkowski spacetime are related to the oblate spheroidal coordinates $(T,r,\theta,\tilde{\Phi})$ by $\tilde \Phi=\Phi+\arctan\frac{a}{r}$.}
\begin{align}\label{eta-kerr}
	\eta=dT^2-dr^2-\Sigma d\theta^2-(r^2+a^2)\sin^2\theta d\Phi^2+2a\sin^2\theta dr d\Phi\,.
\end{align}
 It is clear that if $a=0$, this metric reduces to the Schwarzschild metric \eqref{Schw}. The coordinates $(T,r,\theta,\Phi)$ can be obtained from the standard Boyer---Lindquist coordinates ($t,r,\theta,\varphi)$ by the following transformation:
 \begin{align}\label{KStoBL}
 	T=t+\int^r\frac{2Mr'}{\Delta(r')}dr'\,,\qquad \Phi=\varphi+\int^r\frac{a}{\Delta(r')}dr'\,,
 \end{align}
  where $\Delta=r^2-2Mr+a^2$. Note that the metric \eqref{Kerr-metric} is explicitly nonsingular at the horizons: the coordinate singularities of the Boyer–Lindquist metric are removed by the transformation \eqref{KStoBL}.

Following the procedure described in Section~\ref{sec:idea}, we realize the perturbation using three pairs of embedding functions of the form \eqref{circle-like}. The first two pairs jointly reproduce its dependence on $d(T+r)$, while introducing additional terms involving only $dr,d\theta$ and $d\Phi$. The third pair eliminates the remaining $\Phi\Phi$ component. The residual tensor then depends only on $r,\theta$ and takes the form 
\begin{align}
a^2\left[\left(d\left(\sqrt{\frac{Mr}{2\Sigma}}\cos 2\theta\right)\right)^2+\left(d\sqrt{\frac{3Mr}{2\Sigma}}\right)^2\right]\,,
\end{align}
which admits a straightforward embedding into two dimensions. Combining this with the embedding of the background part, for $a\neq0$ we obtain the following embedding of the Kerr metric:
\begin{align}\label{Kerr-embedding}
\begin{alignedat}{2}
y^0&=T,
&\quad
y^1&=r\cos\theta,
\\
y^2&=(r\sin\Phi+a\cos\Phi)\sin\theta,
&
y^3&=(r\cos\Phi-a\sin\Phi)\sin\theta,
\\[0.5ex]
y^4&=a\sqrt{\dfrac{2Mr}{\Sigma}}\sin\theta
\sin(\frac{T+r}{a}-\Phi),
&
y^5&=a\sqrt{\dfrac{2Mr}{\Sigma}}\sin\theta
\cos(\frac{T+r}{a}-\Phi),
\\
y^6&=a \sqrt{\dfrac{2Mr}{\Sigma}}\cos\theta
\cos(\frac{T+r}{a}),
&
y^7&=a \sqrt{\dfrac{2Mr}{\Sigma}}\cos\theta
\sin(\frac{T+r}{a}),
\\
y^8&=a\sqrt{\dfrac{Mr}{2\Sigma}}\sin2\theta\cos\Phi,
&
y^9&=a\sqrt{\dfrac{Mr}{2\Sigma}}\sin2\theta\sin\Phi,
\\
y^{10}&=a\sqrt{\dfrac{Mr}{2\Sigma}}\cos2\theta,
&
y^{11}&=a\sqrt{\dfrac{3Mr}{2\Sigma}}
\end{alignedat}
\end{align}
  in a flat spacetime with signature $(+ - - - - - - -  + + + +)$. Notice that the formulas for $y^0,y^1,y^2,y^3$ can be interpreted as the transformation from the coordinates \eqref{eta-kerr} to the standard Minkowski coordinates. This embedding has the following properties:
    \begin{enumerate}
  	\item It is asymptotically flat: the embedding functions $y^4\ldots y^{11}$ tend to zero as $r\to\infty$ at the same rate.
  	\item It has no Schwarzschild limit: as $a\to 0$ it becomes an embedding of Minkowski spacetime into itself. 
  	\item  It covers one asymptotically flat exterior region and the associated black-hole interior down to $r=0$, extending smoothly across both the outer and inner horizons. Note that this embedding does not cover the maximally extended Kerr spacetime in its entirety.
    
  	\item The embedding of the equatorial submanifold $\theta=\pi/2$ into a six-dimensional flat space with signature $(++----)$ takes an exceptionally simple form after an orthogonal rotation, removal of the identically vanishing coordinates, and relabeling of the remaining embedding functions in \eqref{Kerr-embedding}:
\begin{align}
\begin{alignedat}{2}
y^1&=T,
&\qquad
y^2&=a\sqrt{\frac{2M}{r}},
\\[1ex]
y^3&=r\cos\Phi-a\sin\Phi,
&\qquad
y^4&=r\sin\Phi+a\cos\Phi,
\\[1ex]
y^5&=a\sqrt{\frac{2M}{r}}\sin\left(\frac{T+r}{a}-\Phi\right),
&\qquad
y^6&=a\sqrt{\frac{2M}{r}}\cos\left(\frac{T+r}{a}-\Phi\right).
\end{alignedat}
\end{align}
To the best of our knowledge, the lowest-dimensional previously known explicit isometric embedding of the three-dimensional equatorial Kerr submanifold was constructed in \cite{Hong:2012fa} and required $12$ ambient dimensions.
      	\item This embedding can be generalized to an embedding of the Kerr-Newman-AdS black hole that smoothly extends across all horizons. We will construct it in the next section.
  \end{enumerate}
Finally, the remarkably simple form of the embedding \eqref{Kerr-embedding} suggests that it may admit an algebraic description. In Appendix~\ref{App:Algebraic}, we show that its image lies in an auxiliary five-dimensional semialgebraic set, much like a helix on a cylinder.
\section{Embedding of any metric of Kerr-Schild-like form}
The form of the embedding \eqref{Kerr-embedding} could make one wonder whether it is possible to embed a wider class of metrics using a similar ansatz. This is indeed possible. Suppose that the perturbation $h$ in $g=\bar g+h$ has the form
\begin{align}\label{h}
	h=\epsilon f^2(d\psi-pd\phi)^2\,,
\end{align}
where $f$, $\psi$, $p$, $\phi\in C^\infty(M)$, $0\leq p\leq 1$ and $\epsilon=\pm1$. Then the following embedding functions in a flat space with signature $\epsilon(++++----)$
\begin{align}\label{emb-h1}
\begin{alignedat}{2}
y^1&=f\sqrt{p}\sin(\psi-\phi),
&\qquad
y^2&=f\sqrt{p}\cos(\psi-\phi),
\\[1ex]
y^3&=f\sqrt{1-p}\cos\psi,
&\qquad
y^4&=f\sqrt{1-p}\sin\psi,
\\[1ex]
y^5&=f\sqrt{p(1-p)}\cos\phi,
&\qquad
y^6&=f\sqrt{p(1-p)}\sin\phi,
\\[1ex]
y^7&=\frac{f}{2}(2-p),
&\qquad
y^8&=\frac{\sqrt{3}}{2}fp
\end{alignedat}
\end{align}
together with an isometric embedding of the background metric give, wherever the displayed functions are smooth, an isometric embedding of the full metric $g$. Note that there is one subtlety in this embedding: at points where $p=0$ or $p=1$, smoothness may require replacing the nonnegative square roots by suitable smooth signed square roots, whenever these exist. For example, in the Kerr case, we use $\cos\theta$ rather than $|\cos\theta|$ as a signed square root of $1-\sin^2\theta$ to ensure smoothness across the equatorial plane.

The particular case $p=0$ gives 
\begin{align}\label{p=0}
	h=\epsilon f^2(d\psi)^2\,.
\end{align}
In this case, we have the following embedding of $h$ into three-dimensional flat space with the ambient metric $\eta=\epsilon\left((dy^1)^2+(dy^2)^2-(dy^3)^2\right)$:
\begin{align}
	\begin{split}
		y^{1}=f\cos\psi\,\qquad y^2=f\sin\psi\,,\qquad y^3=f\,.
	\end{split}
\end{align}
The Schwarzschild perturbation in \eqref{Schw} belongs to this case, and the embedding constructed in Section~\ref{sec:schw} can be obtained immediately.

 Using \eqref{emb-h1} we can easily construct, for example, an embedding of the Kerr-Newman-AdS spacetime. Indeed, the corresponding metric has the form (see, e.g., \cite{Hassaine:2024mfs})
 \begin{align}\label{Kerr-NA}
     g=\bar g_{\mathrm{AdS}}-\frac{2mr-q^2}{\Sigma}\left(dT-\frac{a\sin^2\theta}{1-a^2/\ell^2}d\Phi+\frac{\Sigma}{(r^2+a^2)(1+r^2/\ell^2)}dr\right)^2\,,
 \end{align}
 where $\bar g_{\mathrm{AdS}}$ is the AdS metric and $\Sigma=r^2+a^2\cos^2\theta$. The perturbation in \eqref{Kerr-NA} belongs to the class of perturbations $h$ defined by \eqref{h} with 
 \begin{align}
	\begin{split}\label{KN-AdS}
		&p=\sin^2\theta\,,\qquad \epsilon=-1\,,\qquad f=\frac{|a|}{1-a^2/\ell^2}\sqrt{\frac{2mr-q^2}{\Sigma}}\,,\\
		&\psi=\frac{1-a^2/\ell^2}{a}\left(T+\ell\arctan\frac{r}{\ell}\right)\,,\qquad \phi=\Phi+\arctan \frac{r}{a}-\frac{a}{\ell}\arctan\frac{r}{\ell}\,
	\end{split}
\end{align}
for $r>q^2/2m$ and $0<|a|<\ell$. Combining this result with an embedding of the AdS spacetime into five dimensions, we obtain a 13-dimensional embedding of the Kerr-Newman-AdS spacetime which covers the region $r>q^2/2m$. Despite this restriction, it smoothly covers all horizons. This construction does not extend to the entire region $r>0$ because the coefficient of the perturbation changes sign at $r=q^2/(2m)$ \footnote{Note, however, that the time periodicity of the standard five-dimensional AdS embedding means that, for unwrapped time, the resulting map is, strictly speaking, an isometric immersion rather than a global embedding.}. Nevertheless, this limitation can be overcome by splitting $h$ into two terms, $h=h_1+h_2$, each with a constant overall sign, and embedding them separately, which requires additional ambient dimensions.

The question naturally arises whether it is possible to write a rank $1$ perturbation $h$ in the form \eqref{h}.  In $4$ dimensions and locally, the answer is almost affirmative. Indeed, the rank condition locally implies $h_{\mu\nu}=Hl_\mu l_\nu$ for some function $H$ and a $1$-form $l$. In $d=4$, we have $l\wedge dl\wedge dl=0$, and hence, by the Pfaff-Darboux theorem \cite{olver1993applications}, on a region of constant Pfaff rank we can write $l$ as
  \begin{align}
 	l=k(d\psi-pd\phi)\,,
 \end{align}
where $k$, $\psi$, $\phi$, $p$ are smooth functions and $k\neq0$. Comparing with \eqref{h}, we choose $f=|k|\sqrt{|H|}$ and $\epsilon=\mathrm{sign}(H)$. If $p=0$ throughout the region, then \eqref{p=0} holds. Otherwise, after a constant shift of $p$ and a corresponding adjustment of $\psi$ if necessary, a suitable local rescaling of $\phi$ and $p$ yields $0<p<1$. Note that the representation $h=\epsilon f^2(d\psi-p\,d\phi)^2$ is not unique. Different choices of $f,\psi,p,\phi$ representing the same tensor $h$ can lead to isometric embeddings with different extrinsic geometries. In particular, this freedom can be used to construct other twelve-dimensional embeddings of the Kerr metric that are not related to \eqref{Kerr-embedding} by isometries of ambient space. We leave this question for future work.

We see that \eqref{emb-h1} provides a constructive formula for the local embedding of anything which belongs to the (generalized) KS class in four dimensions and possibly beyond that. One could push this further and turn this formula into an embedding of almost anything. It can be achieved if one refrains from decomposing  the metric into a "background" $\eta$ and a "perturbation" $h$ and instead presents it as four terms of the form $h$ with different functions $f_\mu$, $p_\mu$, $\psi_\mu$ and $\phi_\mu$. This gives us sixteen scalar functions in total, precisely the number of independent components of a Lorentzian tetrad, so it allows us, in principle, to embed any metric with a known tetrad into a 32-dimensional ambient spacetime. Moreover, using the six-parameter local Lorentz gauge freedom, we could reduce, under some additional regularity conditions, three of the four $p_\mu$ and $\phi_\mu$ to zero (see Theorem 4 in \cite{Llosa_2012}), so three of the four terms $h_\mu$ would have the form $h_i = \epsilon_i f^2_i d\psi_i^2$ (embeddable in 3D), so the total number of required dimensions is $3\cdot3+8=17$ once such a gauge is fixed. However, such gauge fixing for an arbitrary metric is itself a complicated problem.

{\textbf{AI usage disclosure}}. 
ChatGPT was used in preparing parts of this article. In particular, the isometric embedding of the Kerr spacetime \eqref{Kerr-embedding} was first obtained using GPT-5.6 with a prompt explicitly referring to \cite{statja70}.  All results obtained with AI assistance were subsequently verified by the authors, both using Maple and by hand.

 {\textbf{Acknowledgements}}. The authors are grateful to S. A. Paston for useful discussions.  M.M. also acknowledges fruitful discussions with Thomas Basile and Viacheslav Krivorol. The work of M.M. was supported by the ULYSSE Incentive Grant for Mobility in Scientific Research [MISU] F.6003.24, F.R.S.-FNRS, Belgium.
 \appendix
 \section{Algebraic properties of the Kerr embedding}\label{App:Algebraic}
We begin with a simple example illustrating the construction below. Consider the cylinder $C=S^1\times\mathbb R$, viewed as a trivial $U(1)$-bundle over $\mathbb R$, with angular coordinate $\phi\in\mathbb R/(2\pi\mathbb Z)$ and base coordinate $\rho\in\mathbb R$. Its standard realization in $\mathbb R^3$ is algebraic and hence semialgebraic, meaning that it can be described by finite combinations of polynomial equations and inequalities in the ambient coordinates. However, the image of the section $\phi=\rho\pmod{2\pi}$ is a circular helix, which is not semialgebraic. We now show that the Kerr embedding admits an analogous description, with its image playing the role of the helix inside a higher-dimensional semialgebraic manifold.
 
 As in the example above, the surface given by \eqref{Kerr-embedding} is not semialgebraic \textit{per se}.  Indeed, its intersection with the planes defined by $y^1=\mathrm{const}>0$, $y^2=y^3=0$ is a circular helix, which is not semialgebraic. This is incompatible with the semialgebraicity of the original surface.
 
 To construct the counterpart of the cylinder, let us replace the combination $\frac{T+r}{a}$ in the arguments of the trigonometric functions in $y^4$, $y^5$, $y^6$, $y^7$ with an independent auxiliary variable  $\psi\in \mathbb{R}/2\pi\mathbb{Z}$. After this replacement, one can show that the five-dimensional surface $M_5$ defined by the modified \eqref{Kerr-embedding} is indeed semialgebraic and the induced metric on it is
  	\begin{align}\label{ds5}
 	g_5=\eta-\frac{2Mra^2}{\Sigma}
 	\left(d\psi-\sin^2\theta\,d\Phi\right)^2.
 \end{align}
 By construction, $M_5$ is a trivial $U(1)$-bundle with $\psi$ being a fiber coordinate. The preferred section $\psi=\frac{T+r}{a}$ equipped with the induced metric is precisely the Kerr spacetime. We find this analogy between the Kerr embedding and a helix on a cylinder somewhat unexpected.

\end{document}